# Experimental setup and procedure for the measurement of the $^7$Be(n,p)$^7$Li reaction at n_TOF

M. Barbagallo[1], J. Andrzejewski[2], M. Mastromarco[1], J. Perkowski[2], L. A. Damone[1,16], A. Gawlik[2], L. Cosentino[3], P. Finocchiaro[3], E. A. Maugeri[4], A. Mazzone[1,33], R. Dressler[4], S. Heinitz[4], N. Kivel[4], D. Schumann[4], N. Colonna[1], O. Aberle[5], S. Amaducci[29,31], L. Audouin[6], M. Bacak[5,7,15], J. Balibrea[19], F. Bečvář[8], G. Bellia[3], E. Berthoumieux[15], J. Billowes[9], D. Bosnar[10], A. Brown[11], M. Caamaño[12], F. Calviño[13], M. Calviani[5], D. Cano-Ott[19], R. Cardella[5], A. Casanovas[13], F. Cerutti[5], Y.H. Chen[6], E. Chiaveri[5,9,14], G. Cortés[13], M. A. Cortés-Giraldo[14], S. Cristallo[43], M. Diakaki[15], M. Dietz[17], C. Domingo-Pardo[18], E. Dupont[15], I. Durán[12], B. Fernández-Domínguez[12], A. Ferrari[5], P. Ferreira[20], V. Furman[21], K. Göbel[22], A. R. García[19], S. Gilardoni[5], T. Glodariu[23], I. F. Gonçalves[20], E. González-Romero[19], E. Griesmayer[7], C. Guerrero[14], F. Gunsing[15,5], H. Harada[24], J. Heyse[25], D. G. Jenkins[11], E. Jericha[7], K. Johnston[5], F. Käppeler[26], Y. Kadi[5], A. Kalamara[27], P. Kavrigin[7], A. Kimura[24], M. Kokkoris[27], M. Krtička[8], D. Kurtulgil[22], C. Lederer[17], H. Leeb[7], J. Lerendegui-Marco[14], S. Lo Meo[28,29], S. J. Lonsdale[17], D. Macina[5], A. Manna[29,31], J. Marganiec[2,30], T. Martínez[19], J. G. Martins-Correia[5,45], A. Masi[5], C. Massimi[29,31], P. Mastinu[32], E. Mendoza[19], A. Mengoni[28], P. M. Milazzo[34], F. Mingrone[5], A. Musumarra[3,35], A. Negret[23], R. Nolte[30], A. Oprea[23], A. D. Pappalardo[3], N. Patronis[36], A. Pavlik[37], M. Piscopo[3], I. Porras[38], J. Praena[38], J. M. Quesada[14], D. Radeck[30], T. Rauscher[39,40], R. Reifarth[22], M. S. Robles[12], C. Rubbia[5], J. A. Ryan[9], M. Sabaté-Gilarte[5,14], A. Saxena[41], J. Schell[5,44], P. Schillebeeckx[25], P. Sedyshev[21], A. G. Smith[9], N. V. Sosnin[9], A. Stamatopoulos[27], G. Tagliente[1], J. L. Tain[18], A. Tarifeño-Saldivia[13], L. Tassan-Got[6], S. Valenta[8], G. Vannini[29,31], V. Variale[1], P. Vaz[20], A. Ventura[29], V. Vlachoudis[5], R. Vlastou[27], A. Wallner[42], S. Warren[9], C. Weiss[7], P. J. Woods[17], T. Wright[9], P. Žugec[10,5]

The n_TOF Collaboration

1. Istituto Nazionale di Fisica Nucleare, Sezione di Bari, Italy
2. University of Lodz, Poland
3. INFN Laboratori Nazionali del Sud, Catania, Italy
4. Paul Scherrer Institut (PSI), Villingen, Switzerland
5. European Organization for Nuclear Research (CERN), Switzerland
6. Institut de Physique Nucléaire, CNRS-IN2P3, Univ. Paris-Sud, Université Paris-Saclay, F-91406 Orsay Cedex, France
7. Technische Universität Wien, Austria
8. Charles University, Prague, Czech Republic
9. University of Manchester, United Kingdom
10. Department of Physics, Faculty of Science, University of Zagreb, Bijenička c. 32, 10000 Zagreb, Croatia
11. University of York, United Kingdom
12. University of Santiago de Compostela, Spain
13. Universitat Politècnica de Catalunya, Spain
14. Universidad de Sevilla, Spain
15. CEA Irfu, University Paris-Saclay, Gif-sur-Yvette, France
16. Dipartimento di Fisica, Università degli Studi di Bari, Italy
17. School of Physics and Astronomy, University of Edinburgh, United Kingdom
18. Instituto de Física Corpuscular, Universidad de Valencia, Spain
19. Centro de Investigaciones Energéticas Medioambientales y Tecnológicas (CIEMAT), Spain
20. Instituto Superior Técnico, Lisbon, Portugal
21. Joint Institute for Nuclear Research (JINR), Dubna, Russia
22. Goethe University Frankfurt, Germany
23. Horia Hulubei National Institute of Physics and Nuclear Engineering, Romania

24. *Japan Atomic Energy Agency (JAEA), Tokai-mura, Japan*
25. *European Commission, Joint Research Centre, Geel, Retieseweg 111, B-2440 Geel, Belgium*
26. *Karlsruhe Institute of Technology, Campus North, IKP, 76021 Karlsruhe, Germany*
27. *National Technical University of Athens, Greece*
28. *Agenzia nazionale per le nuove tecnologie (ENEA), Bologna, Italy*
29. *Istituto Nazionale di Fisica Nucleare, Sezione di Bologna, Italy*
30. *Physikalisch-Technische Bundesanstalt (PTB), Bundesallee 100, 38116 Braunschweig, Germany*
31. *Dipartimento di Fisica e Astronomia, Università di Bologna, Italy*
32. *Istituto Nazionale di Fisica Nucleare, Sezione di Legnaro, Italy*
33. *Consiglio Nazionale delle Ricerche, Bari, Italy*
34. *Istituto Nazionale di Fisica Nucleare, Sezione di Trieste, Italy*
35. *Dipartimento di Fisica e Astronomia, Università di Catania, Italy*
36. *University of Ioannina, Greece*
37. *University of Vienna, Faculty of Physics, Vienna, Austria*
38. *University of Granada, Spain*
39. *Department of Physics, University of Basel, Switzerland*
40. *Centre for Astrophysics Research, University of Hertfordshire, United Kingdom*
41. *Bhabha Atomic Research Centre (BARC), India*
42. *Australian National University, Canberra, Australia*
43. *INAF – Osservatorio Astronomico di Collurania, Teramo, and INFN – Sezione di Perugia, Perugia, Italy*
44. *Institute for Materials Science and Center for Nanointegration Duisburg-Essen (CENIDE), University of Duisburg-Essen, 45141 Essen, Germany*
45. *C2TN, Centro de Ciências e Tecnologias Nucleares, Instituto Superior Técnico, Universidade de Lisboa, Portugal*

**Abstract**

Following the completion of the second neutron beam line and the related experimental area (EAR2) at the n_TOF spallation neutron source at CERN, several experiments were planned and performed. The high instantaneous neutron flux available in EAR2 allows to investigate neutron indiced reactions with charged particles in the exit channel even employing targets made out of small amounts of short-lived radioactive isotopes. After the successful measurement of the $^7$Be(n,α)α cross section, the $^7$Be(n,p)$^7$Li reaction was studied in order to provide still missing cross section data of relevance for Big Bang Nucleosynthesis (BBN), in an attempt to find a solution to the cosmological Lithium abundance problem. This paper describes the experimental setup employed in such a measurement and its characterization.

## 1 Introduction

In July 2014 the second experimental area (EAR2) of the n_TOF spallation neutron-time-of-flight facility at CERN came into operation. The advantage of n_TOF, with respect to other neutron facilities in the world, is the extremely high instantaneous neutron flux delivered in a short time interval at the sample position. The still higher neutron flux of about $10^7 \div 10^8$ n/cm$^2$/s, obtained with the reduced flight path of 19 m with respect to the 185 m of the older EAR1, allows to perform experiments on low-mass targets and/or targets made out of short-lived radionuclides, even on isotopes characterized by a small reaction cross-section, with a favorable signal to background ratio. Indeed, challenging measurements of reactions with outgoing charged particles have now become attainable [1][2][3].

The first experiment done at EAR2 was the measurement of the energy-dependent $^7$Be(n,α)α cross-section [4],[5], of relevance for a possible nuclear solution to the cosmological Lithium abundance problem (CLIP) in the Big Bang Nucleosynthesis (BBN) framework [6],[7]. Before the n_TOF measurement, the only existing data consisted in a single value measured in the 1960s at thermal neutron energy [8]. The new measurement indicated that the cross section of the (n,α) reaction was too low to significantly affect the abundance of primordial Lithium. Following that measurement, the last piece of information still missing in the BBN and CLIP scenario was an accurate measurement of the $^7$Be(n,p)$^7$Li cross-section. Indeed, the only two existing measurements on this reaction in the neutron energy range from thermal to keV date back to the late 1980s and are in disagreement with each other [9],[10]. Moreover, above approximately 10 keV the cross-section must match the data available from the time-reversal reaction [11], a check that could not be performed with the old data, as they stopped at neutron energies well below those of the $^7$Li(p,n)$^7$Be data. The neutron flux features of the EAR2 facility and the n_TOF time-energy dynamic range provided the opportunity of a high quality direct measurement of the $^7$Be(n,p)$^7$Li cross-section.

Contrary to $^7$Be(n,α)α reaction, with no need for a pure $^7$Be target because of the signature consisting in two ≈8 MeV α-particles, in the present case an isotopically pure $^7$Be target was mandatory. The $^7$Be(n,p)$^7$Li reaction has a very high thermal cross-section (several $10^4$ b), but it produces rather low-energy protons of 1.44 MeV which could easily interfere with background from other reaction channels on the sample backing or contaminants. The main requirements were thus to run the experiment on an as much as possible pure target and to detect the emitted proton in a very selective fashion.

This paper describes the experimental setup and the validation test, performed on a $^6$LiF target, which allowed us to prove the feasibility of the experiment and to provide an absolute reference to normalize the cross-section. The reaction on the $^7$Be target was measured in the energy range from thermal to ≈400 keV, and some preliminary data are shown. The detailed physics analysis is currently being finalized, and the results are going to be published soon [12].

## 2 Experimental setup

Contrary to the measurement of the $^7$Be(n,α) reaction, the high cross-section of the (n,p) channel allowed for a lower efficiency detection system that could be placed off-beam. The main advantages of such a setup were the reduction of: (i) the pile-up issues; (ii) the background of 478 keV γ-rays following the natural decay of the $^7$Be target nuclei into $^7$Li; and (iii) the huge background due to the so-called γ-flash, i.e. the big prompt burst of γ-rays and relativistic charged particles produced by the n_TOF spallation target. These three issues would have posed severe limitations on the performance of an in-beam detector arrangement similar to the one exploited in refs.[4],[5].

In order to use the identification of the emitted 1.44 MeV protons as signature of the reaction under study, a silicon telescope detector was chosen. Besides providing an absolute energy response, silicon has a low sensitivity to γ-rays and scattered neutrons. Indeed, γ-rays and neutrons up to ≈1 MeV basically produce low amplitude signals, with an interaction probability $< 10^{-3} \div 10^{-4}$. Furthermore, they can practically interact only with one of the elements of a telescope, and are therefore efficiently rejected by the coincidence technique. To maximize the geometrical efficiency a reasonably wide detector was needed, and for this reason a 5cm x 5cm geometry was chosen. Finally, in order to circumvent the noise problem associated with large area silicon detectors, in particular for the thinner ΔE stage characterized by a large capacitance, it was decided to employ a strip-like geometry. Both ΔE and E stages were made of 16 strips 3 mm wide and 50 mm long, with an inter-strip gap of 0.1 mm. The thickness was respectively 20 and 300 μm, with a 7 mm distance between the two stages. The nominal energy loss of a 1.4 MeV proton impinging perpendicularly on the telescope is 0.86 MeV on the ΔE stage before it is stopped on the E stage. Additional benefits of the strip detectors are: (i) rejection of spurious background by means of a more selective geometrical choice of the ΔE-E coincidences; (ii) a better tool to evaluate the geometrical efficiency and the target alignment from the data, which can be easily compared with numerical simulations;

(iii) the possibility of roughly checking the expected isotropy of the angular distribution of the emitted protons.

In Figure 1 a 3D sketch and a vertical profile of the detector setup are shown, not to scale and with exaggerated thickness of the E and ΔE silicon detectors for better clarity. Figure 2 shows a picture of the real setup on the bench before its installation on the EAR2 neutron beam line. In Figure 3 we show the scattering chamber installed on the vertical neutron beam line in EAR2. The front-end electronics boxes are visible, whereas the target and the telescope detector are inside.

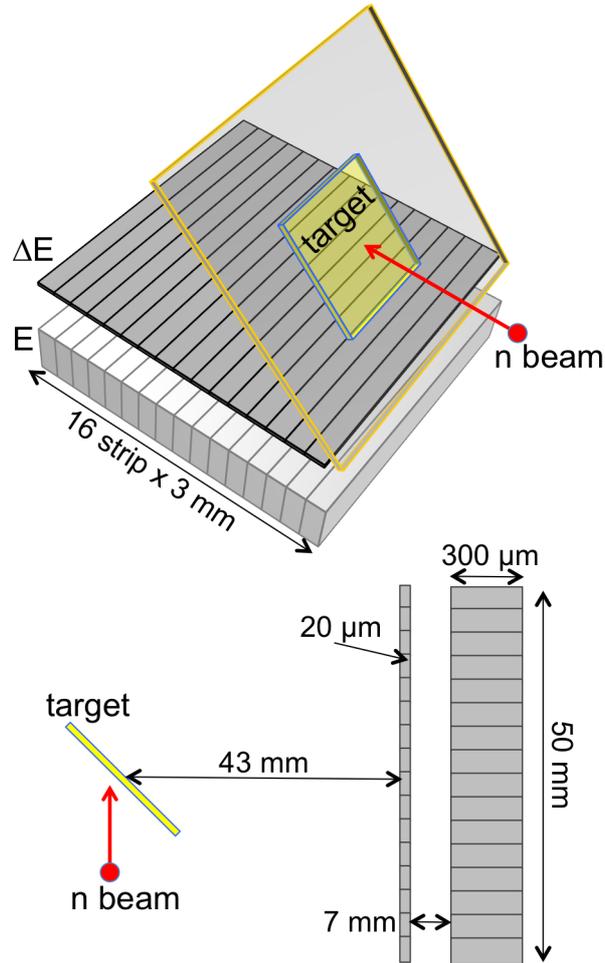

Figure 1. 3D sketch (top) and vertical profile (bottom) of the detector setup. The drawings are not to scale, and the ΔE and E thicknesses are exaggerated for clarity.

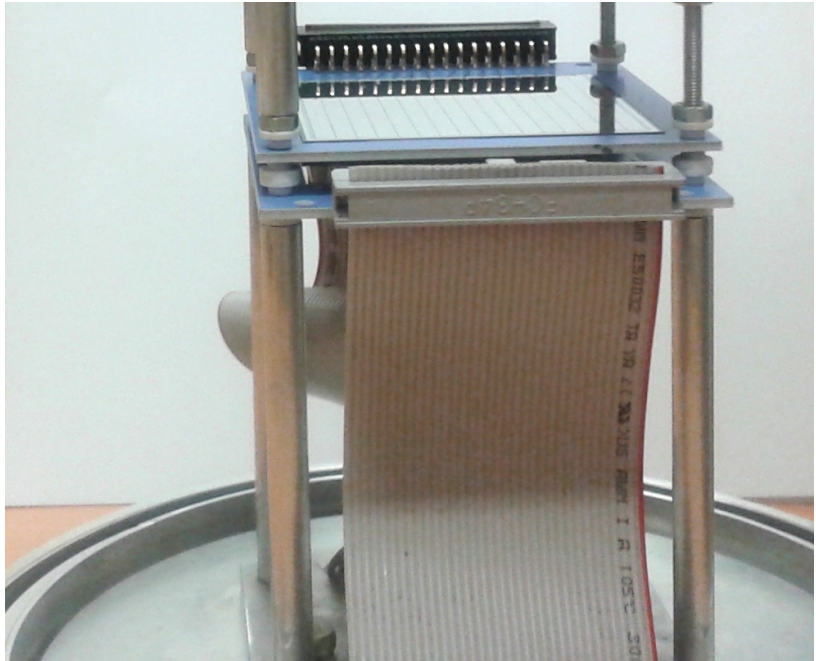

Figure 2. The detector setup on the bench before installation.

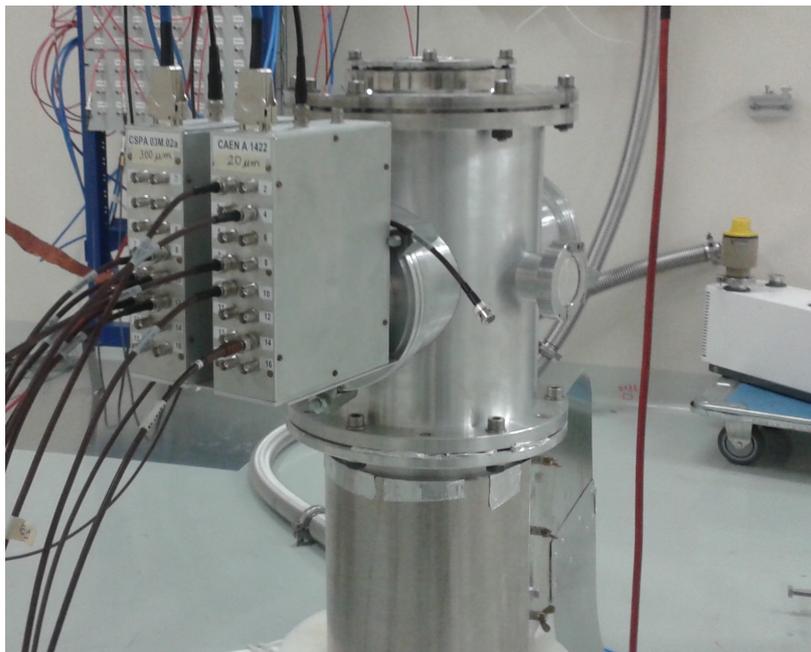

Figure 3. The scattering chamber, with the front-end electronics, installed on the vertical neutron beam line in EAR2. The target and the telescope detector are inside.

A standard commercial front-end and readout electronics was chosen for the setup. The preamplifiers consisted of 16 units of CAEN-A1422H-F3 [13] for the ΔE stage (90 mV/MeV gain), and 16 units of CSPA 03M.02a [14] for the E stage (50 mV/MeV gain). The rise time of the signals from the preamplifiers for the ΔE and E stages were respectively 100 and 50 ns. Two 16-channel CAEN-N568B modules were used as shaping amplifiers, and their shaping time was optimized at 0.2 μs by means of calibration α-sources. The analog signals were then digitized using Acqiris flash ADCs with up to 14 bit resolution and up to 2 GHz sampling rate.

The ΔE and E silicon detectors, along with the front-end electronics, were tested on the bench by means of α-sources. The behavior was quite uniform strip-by-strip, as can be seen in Figure 4 showing the FWHM resolution for each ΔE and E detector strip, tested with a pulser and with $^{238}$U (for the ΔE detector) and $^{241}$Am (for the E detector).

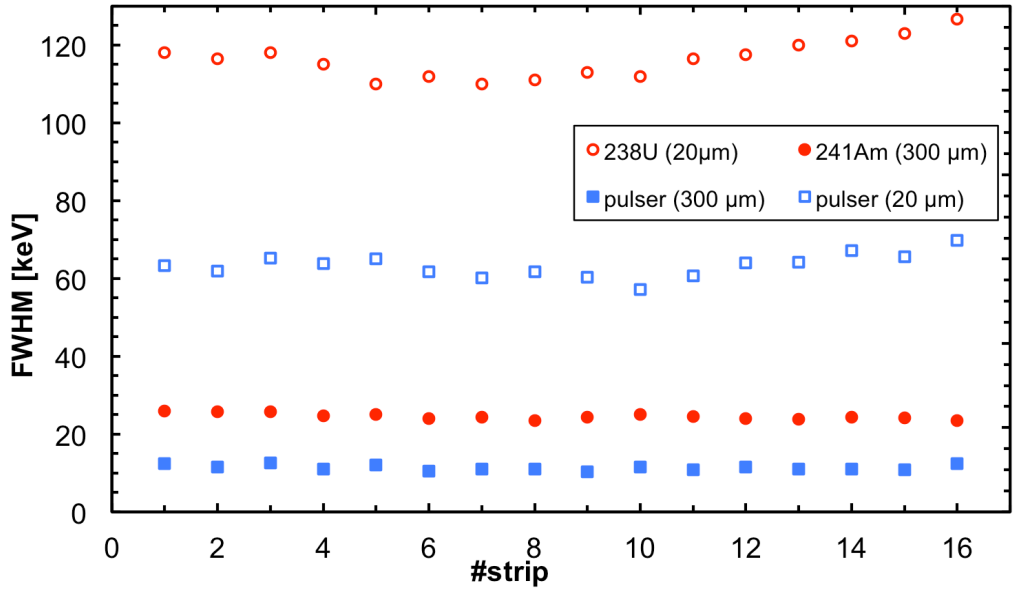

Figure 4. FWHM resolution for each ΔE and E detector strip, tested with a pulser and with α particles from $^{238}$U (for the ΔE detector) and $^{241}$Am (for the E detector).

The $^7$Be target preparation was basically done in two separate steps. First, 200 GBq of $^7$Be were extracted from the cooling water of the SINQ spallation source at PSI and deposited onto a suitable support in form of $^7$Be(NO$_3$)$_2$ colloid. Then, the support was transported to the ISOLDE facility at CERN, where it was installed in the ion source ("ISOLDE oven") to produce a 30 keV ion beam. The $^7$Be beam was separated by means of a magnetic dipole, and was implanted on a 20 μm thick aluminum backing placed in a high vacuum collection chamber. The beam direction was swept during the implantation with the aim of depositing a uniform film of 1.5 cm x 1.5 cm area. However, an accurate measurement of the spatial distribution of the $^7$Be activity in the sample, performed after the measurement, revealed that the isotope had been implanted over a smaller area, showing a Gaussian-like profile of 5 mm FWHM.

Two $^7$Be samples were produced, a test sample with 20 MBq activity and the final sample with 1.1 GBq. A detailed description of the complex production procedure, of the sample characterization and of the target installation will be presented in a forthcoming paper currently in preparation [15].

An additional sample, to be used for the validation test described below, was produced by evaporating a 1.8 μm thick $^6$LiF layer onto a 2 μm thick mylar foil [16]. The size of this target was 5 cm x 5 cm, but a 1.5 cm x 1.5 cm mask was used during the measurement resulting in an effective target area consistent with the planned area of the $^7$Be sample. The target installation required a separate support to be inserted independently, as shown in Figure 5, due to its high activity. The telescope detector had already been installed from the opposite side of the chamber.

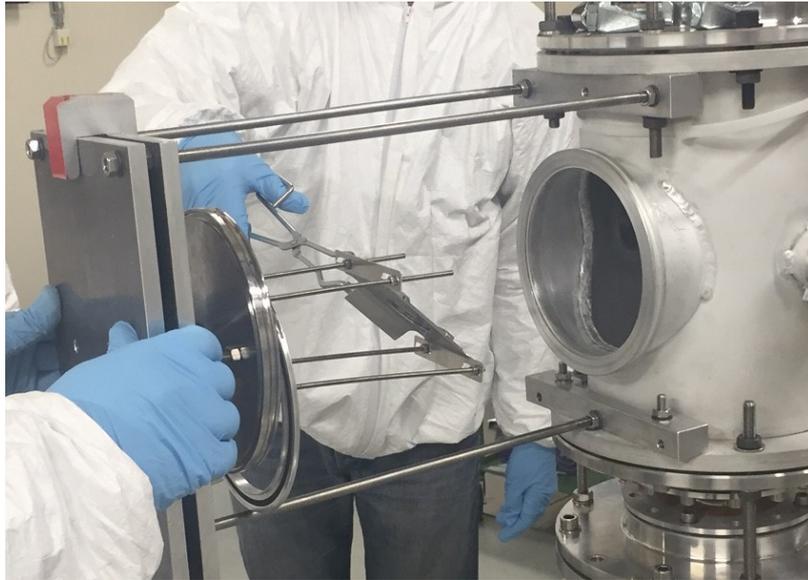

Figure 5. Insertion of the $^7$Be radioactive target in the setup, with a support independent of the telescope detector installed from the opposite side of the chamber.

The detection efficiency of the setup was simulated by means of the Monte Carlo code GEANT4, assuming isotropic emission from the target (the selected physics list was FTFP_INCLXX_HP and emstandard_opt0, as already used in previous works about n_TOF neutron flux **Error! Reference source not found.** and neutron detection **Error! Reference source not found.**). Many different neutron beam energies were simulated, assuming a gaussian beam shape with a sigma value of 1 cm and a radius of 2 cm. The result, in relative units as a function of the strip number, is shown in Figure 6 and Figure 7 for tritons and protons (additional details are provided in sections 3 and 4). The shape of the efficiency accounts for the 45° tilt of the target with respect to the detector.

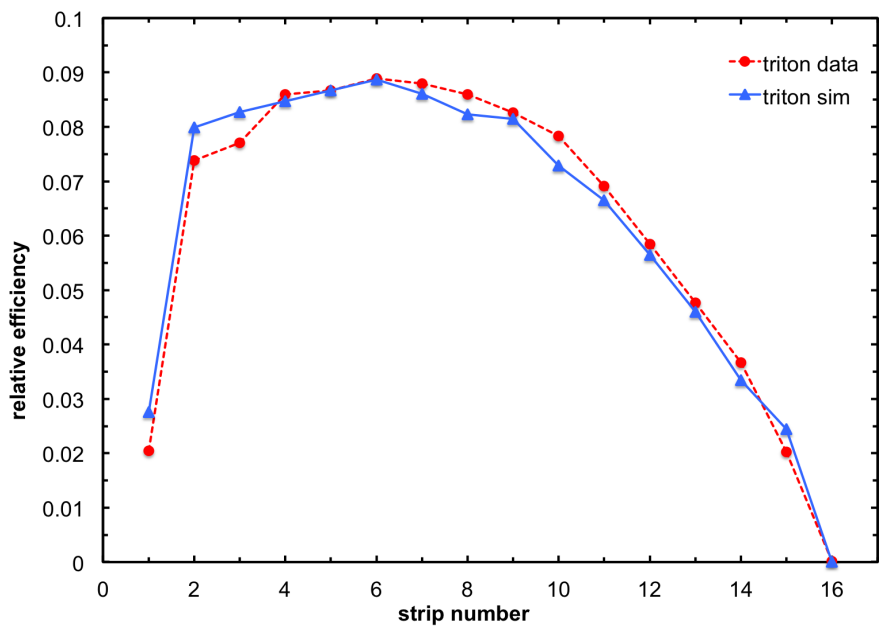

Figure 6. Relative detection efficiency of the setup, as a function of strip number, obtained in the GEANT4 simulation (isotropic emission was assumed) and with $^6$LiF target (triton detection in the validation test).

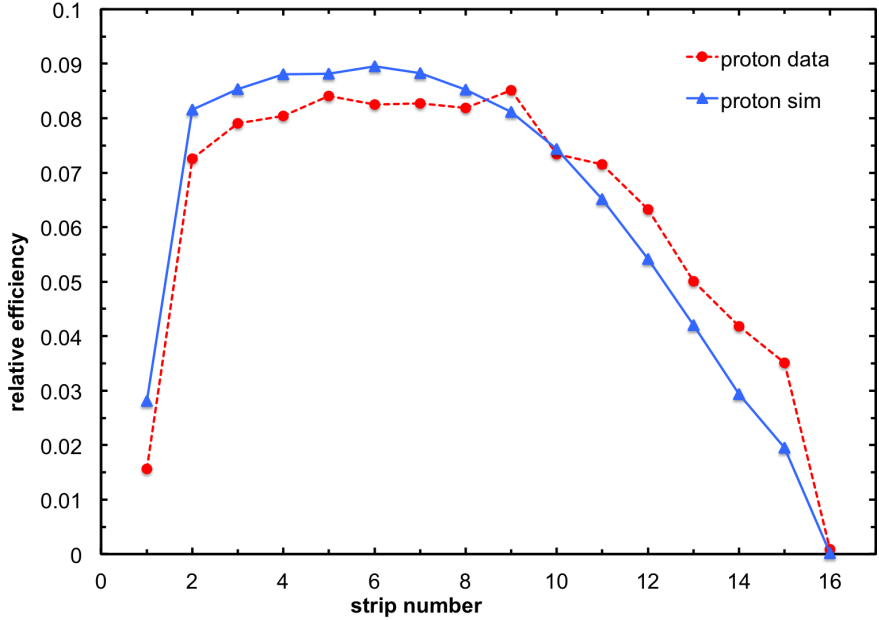

Figure 7. Relative detection efficiency of the setup, as a function of strip number, obtained in the GEANT4 simulation (isotropic emission was assumed) and with $^7$Be target (proton detection).

## 3 The validation test

In order to assess the feasibility of the measurement the $^6$LiF sample was inserted in the setup on the EAR2 beam line. The exploited test reaction was the well known

$$^6Li + n \rightarrow t\,(2.73\ MeV) + \alpha\,(2.05\ MeV) \qquad (1)$$

The α particles were stopped in the ΔE detector layer, therefore they could not be registered in coincidence mode. Conversely, the tritons crossed the ΔE and were stopped in the E layer, releasing respectively 1 and 1.7 MeV (nominal values when tritons are emitted from the very front face of the sample and impinge perpendicularly on the detector). Figure 8 shows the ΔE-E scatter plot measured in the test reaction (1). The triton pattern is clearly visible and its wide energy spread in both directions is due to variable energy loss in the target, depending on the emission depth and angle, and to the spread of incidence angles on the detector.

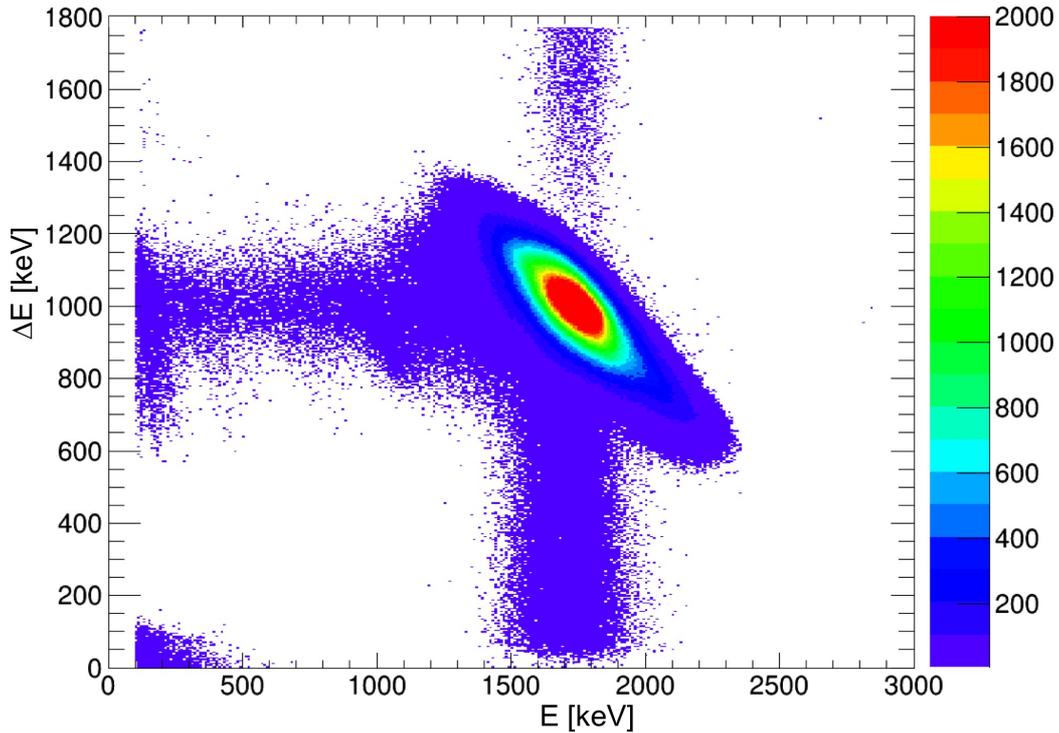

Figure 8. ΔE-E scatter plot measured from the $^6$Li(n,α) reaction, with the tritons locus clearly visible. The wide energy spread was due to the emission depth and angle in the target, and to the wide range of incidence angles on the detector.

By selecting the events falling within the triton locus, and profiting by the information about the hit strip, one can evaluate the relative detection efficiency for the tritons produced in the $^6$Li(n,α) reaction as a function of strip number. The related plot (in Figure 6) shows the measured distribution for low-energy neutrons ($E_n$ < 50 keV), as compared with a GEANT4 simulation assuming isotropic emission from the target (the triton emission in the $^6$Li(n,α) reaction is isotropic for neutron energies below 50 keV). In light of this agreement between data and simulation, the absolute detection efficiency as a function of the incident neutron energy was evaluated by means of a GEANT4 simulation. The results are shown in Figure 9 (continuous line, right-hand axis).

The $^6$Li(n,α) data represent a very reliable absolute reference to normalize the $^7$Be(n,p)$^7$Li data, as the cross-section is an international standard from thermal up to 1 MeV neutron energy [19]. In order to prove their reliability, these data were also used to explicitly reconstruct the $^6$Li(n,α) cross-section, by making use of the known neutron flux in EAR2 which had been determined by means of a set of independent measurements based on different reference reactions and employing several different detector technologies [20]. The good agreement between the so obtained cross-section and the standard one reported in the ENDF database [21] is shown in Figure 9, where the dip in the detection efficiency corresponds to the well-known resonance in the $^6$Li(n,α) cross-section which has a mostly p-wave forward-backward distribution [22]. A final redundant check was also performed with respect to the independent neutron beam monitor SIMON2, which is permanently installed in the beam line and which is based on another $^6$LiF target coupled to four silicon detectors [22]. Also in this case a good agreement was obtained within the experimental uncertainties.

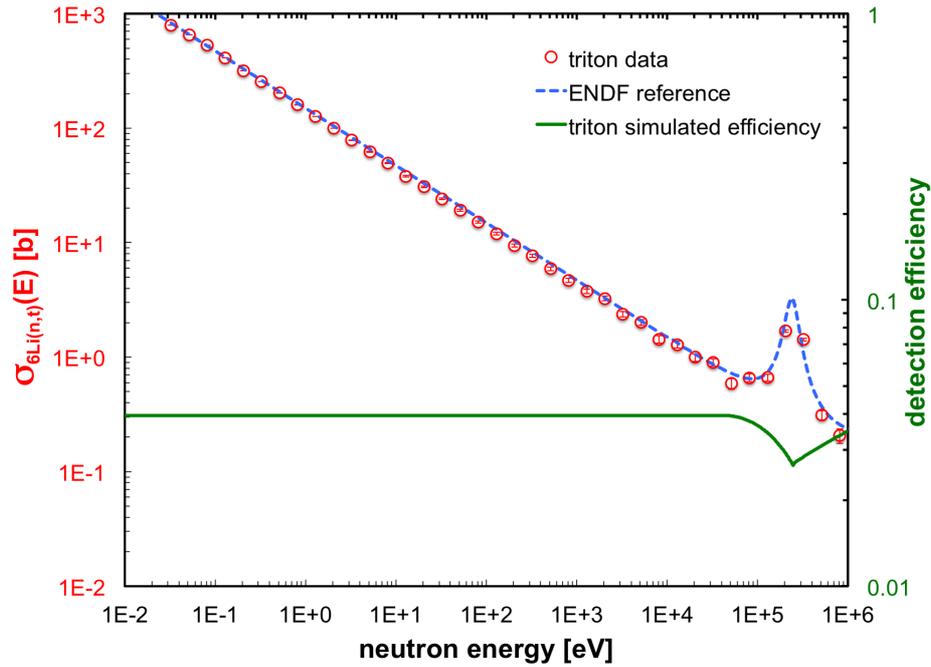

Figure 9. The $^6$Li(n,α)t cross-section (circles), as measured during the validation test, is in good agreement with the international standard (dashed line) and thus can be used for normalization. The green continuous line, to be read on the right-hand axis, represents the triton detection efficiency as simulated by means of GEANT4.

## 4   Preliminary data

A preliminary test run was done with the lower activity $^7$Be target (20 MBq), to further prove the overall feasibility of the measurement. The test was followed by the final measurement with the 1.1 GBq activity target, and both results were in agreement within the statistical uncertainties. In Figure 10 we show the distribution of the time interval between corresponding ΔE and E strips in coincidence events as a function of the neutron time-of-flight. The higher statistics region around 5÷8 ms corresponds to the thermal neutron energy range. Figure 11 shows the projection of Figure 10 onto the Y-axis, that represents the overall coincidence time distribution. One can see that it is centered around 100 ns, due to delays introduced by electronics and cabling, with a FWHM resolution around 23 ns. The two signals from a coincidence, produced by a proton crossing the ΔE detector and stopped in the E detector, are shown in Figure 12.

The width of the time coincidence window between the strips in the ΔE detector and the corresponding strips in the E detector was chosen as 100 ns. Signals on corresponding ΔE and E strips within such a time window were assumed to be proton candidate events. For such events a ΔE-E scatter plot was built, and the geometrical locus ascribed to protons is clearly visible (Figure 13). By selecting the events falling within the proton locus, and profiting by the information about the hit strip, the relative detection efficiency for the protons produced in the $^7$Be(n,p)$^7$Li reaction was evaluated as a function of strip number. The related plot is reported in Figure 7 and compared to the simulation results, where isotropic emission from the target was assumed.

A run with a dummy target, consisting of an aluminum backing without $^7$Be, was performed to evaluate the background contribution due to the target support as a function of neutron energy. The corresponding scatter plot in Figure 14 demonstrates the very low background level achieved by the ΔE-E coincidence technique. The signal-to-background ratio, normalized to the same number of incident neutrons, was plotted in Figure 15. The signal is represented by the number of proton events measured in the $^7$Be(n,p)$^7$Li reaction, and the background is the corresponding number measured with the dummy target. Even up to several 100 keV the signal-to-background ratio remains of the order of 10, thus implying that the cross section under investigation could be measured up to this energy range.

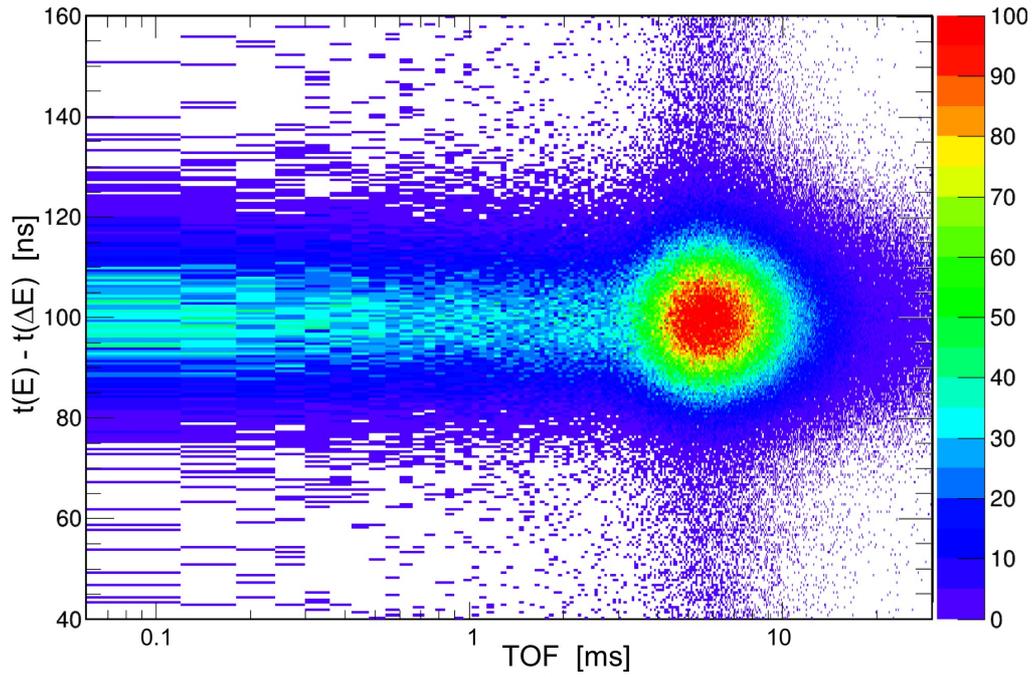

Figure 10. Scatter plot of the time interval between corresponding ΔE and E strips in coincidence events as a function of the neutron time-of-flight, as measured in the $^7$Be(n,p)$^7$Li reaction.

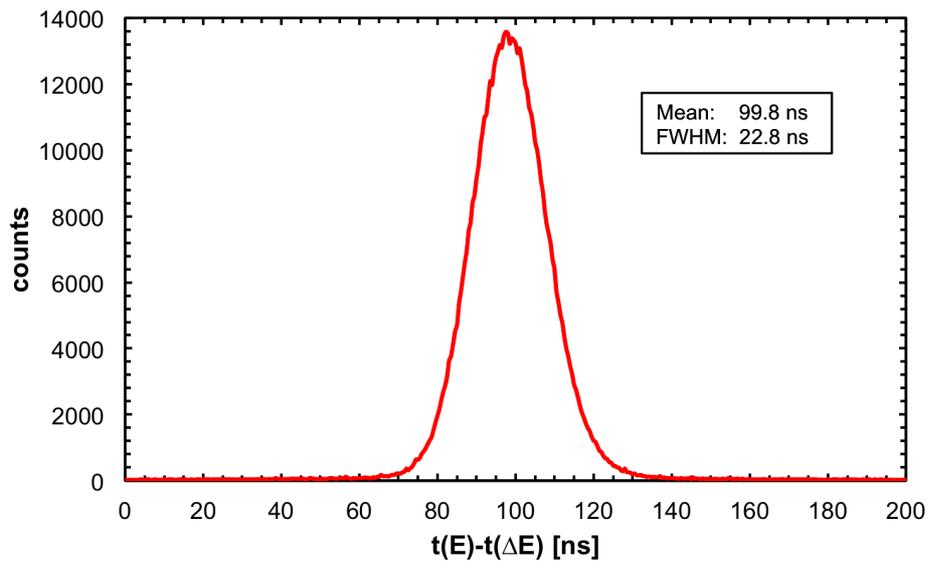

Figure 11. Overall distribution of the coincidence time between corresponding ΔE and E strips, obtained by projecting the scatter plot of Figure 10 on the Y-axis.

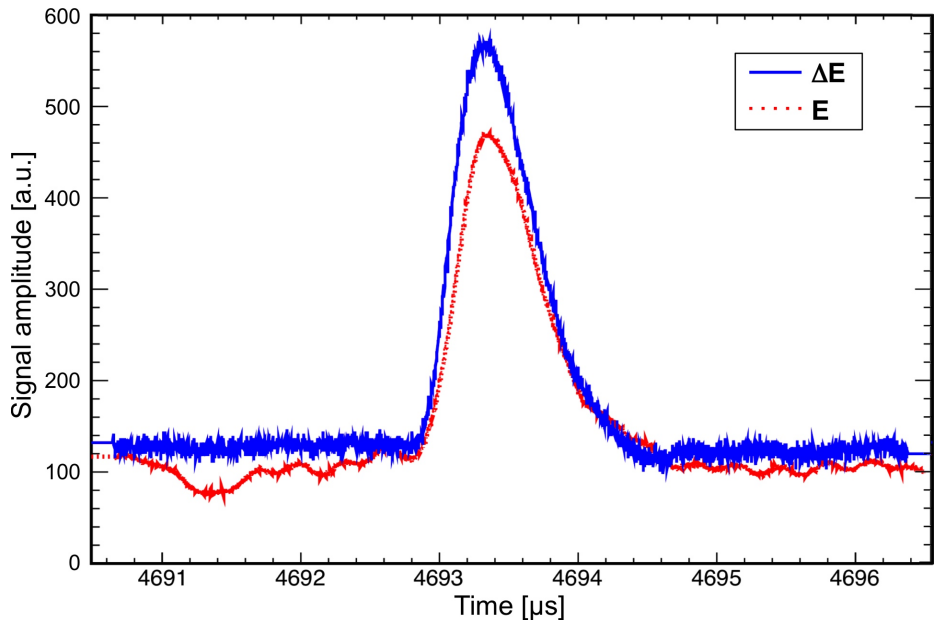

Figure 12. Snapshot of the ΔE and E signals produced by a proton in a strip-strip coincidence event.

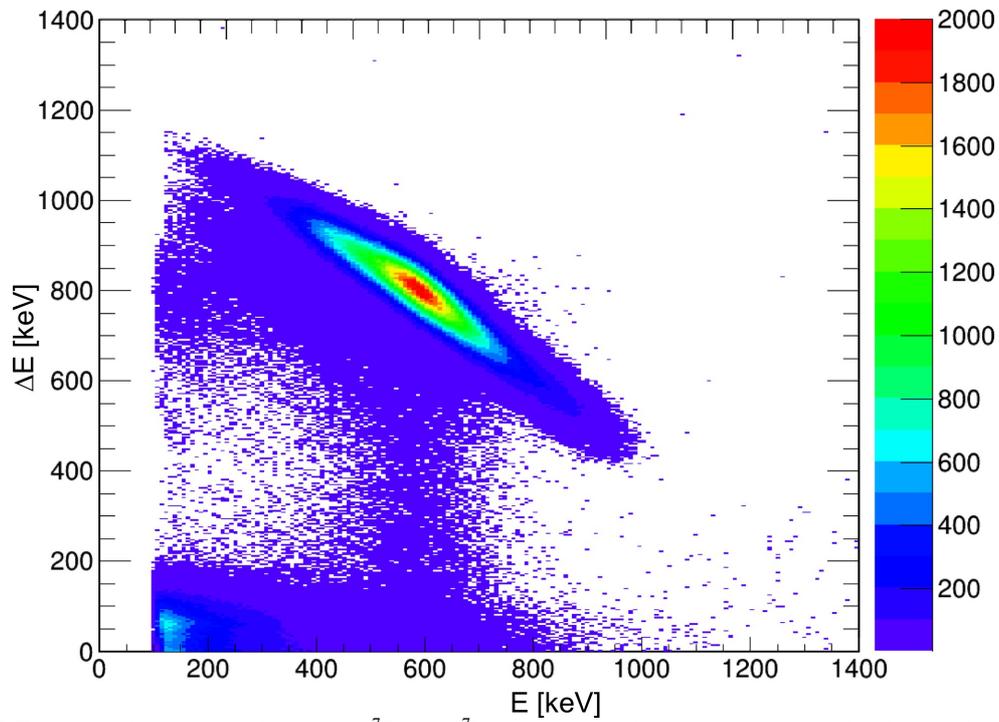

Figure 13. ΔE-E scatter plot measured from the $^7$Be(n,p)$^7$Li reaction, with the proton locus clearly visible.

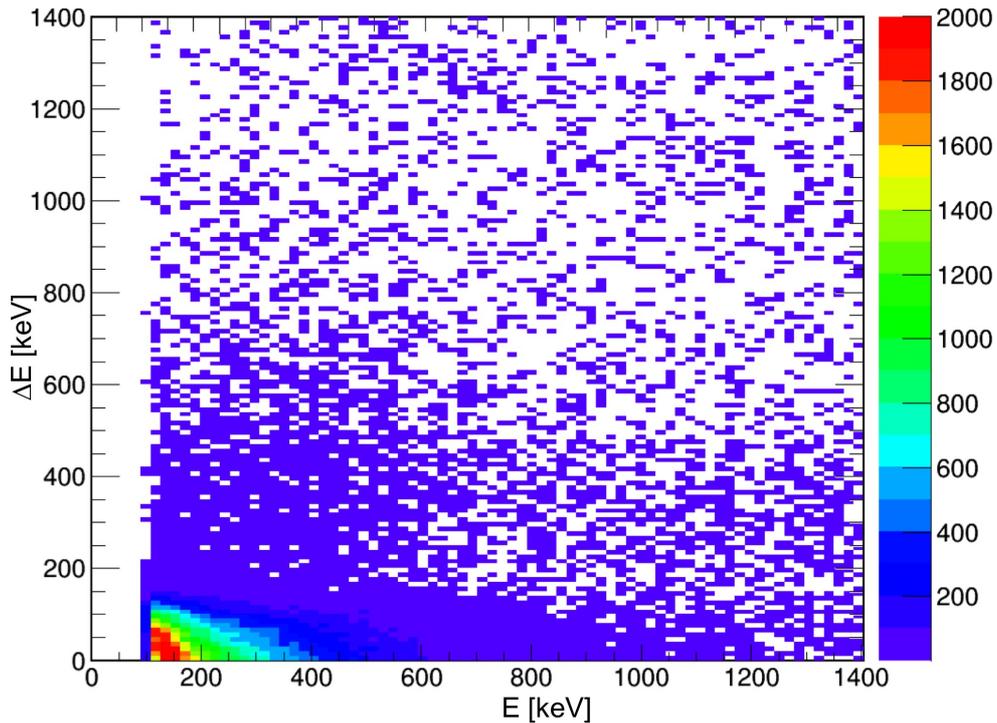

Figure 14. ΔE-E scatter plot measured with a dummy target consisting of the aluminum backing alone.

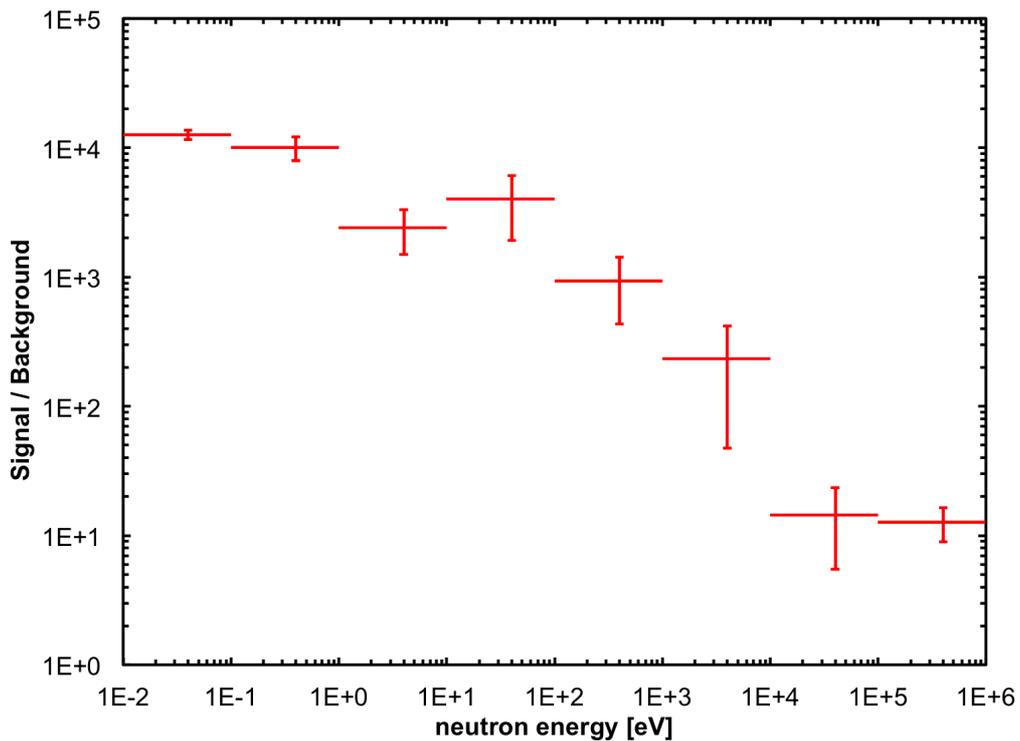

Figure 15. The signal-to-background ratio in eight neutron energy bins. The signal was the number of proton events in the $^7$Be(n,p)$^7$Li reaction. The background data were produced using a dummy target with only the aluminum backing.

## 5   Conclusion

In order to fit one last missing piece of information into the BBN scenario, an accurate measurement of the $^7$Be(n,p)$^7$Li cross-section was planned and performed, as the only two existing measurements from thermal to keV energies on this reaction date back to the 1980s and are in disagreement with each other. The experimental setup described in this paper proved to be reliable and perfectly suited for the task, fulfilling all the expectations and featuring an outstanding signal-to-background performance in a wide energy range, including the one of interest for BBN (i.e. 20÷100 keV). The absolute normalization to the $^6$Li(n,$\alpha$) international standard cross-section, and the redundant normalization to the known neutron flux in EAR2, strongly support the reliability of the resulting cross-section in the neutron energy range from thermal up to ≈400 keV, whose detailed data analysis and physical implications will be discussed in a separate forthcoming paper [12].

# 6 Acknowledgments

The authors are grateful to Carmelo Marchetta (INFN-LNS) for the production of the $^6$LiF test target.

This research was partially funded by the European Atomic Energy Community (Euratom) Seventh Framework Programme FP7/2007-2011 under the Project CHANDA (Grant No. 605203). We acknowledge the support of the Narodowe Centrum Nauki (NCN) under the grant UMO-2012/04/M/ST2/00700-UMO-2016/22/M/ST2/00183, and of the Croatian Science Foundation under project HRZZ 1680.